\documentstyle[aps,epsfig]{revtex}

\draft
\begin{document}
\twocolumn[\hsize\textwidth\columnwidth\hsize
     \csname @twocolumnfalse\endcsname


\title{Tip effects in scanning tunnelling microscopy of atomic-scale
       magnetic structures.
       }

\author{W. A. Hofer$^{*}$ and A. J. Fisher}
\address{
         Department of Physics and Astronomy, University College
         London, Gower Street, London WC1E 6BT, United Kingdom
         }


\maketitle

\begin{abstract}
The spin-polarized scanning tunnelling microscope (STM) can in
principle resolve not only the electronic, but also the magnetic
surface structure. We model recent STM measurements achieving
magnetic resolution on the atomic scale by a first-principles
method. It is shown that the signature of a specific magnetic or
non-magnetic STM tip can unambiguously be identified. It is also
established that the model of Tersoff and Hamann would yield an
electronic as well as magnetic contrast of the surface which is
well below STM resolution. \vspace{0.5 cm}

\end{abstract}

\vskip2pc]

* corresponding author: w.hofer@ucl.ac.uk


Currently, intensive research is undertaken to understand the
complex behavior of spin-systems in thin ferromagnetic (FM) and
antiferromagnetic (AFM) metal films. The motivation behind this
effort is the ever increasing demand for denser storage of
information on magnetic disks, the main storage devices in the
computer industry. While traditionally the emphasis was on long
range spin-interactions, because these can be influenced by
controlled growth of magnetic and non-magnetic multilayers, short
range interactions have only recently come into focus. In this
range neighboring atoms with different magnetic properties can act
as nanomagnets, their magnetic orientation driven by competing
exchange interactions between the electrons. The field gained a
considerable boost last year by the first demonstration of
magnetism on the atomic scale \cite{heinze00}. It is expected that
the packing density of information on ultrathin magnetic films can
ultimately be brought down to the very level of single atoms. We
may therefore not exaggerate by saying that nanomagnetism is today
the most promising field within the burgeoning area of
nanotechnology.

The spin-polarized (SP) scanning tunnelling microscope (STM),
which was used to perform the first local measurements of
antiferromagnetism on an atomic scale, can also serve as a tool to
probe complex spin structures on surfaces, as recently
demonstrated \cite{wortmann01}. In this case a magnetic tip tunes
in on the spin states of electrons on the sample surface and
reveals the magnetic, rather than the electronic, structure of the
sample. But the current theoretical model, employed to elucidate
the magnetic contrast, disagrees with experiments by one order of
magnitude. In order to exploit the full potential of this method
for quantitative measurements of magnetism on the atomic scale it
is necessary to understand in detail how this contrast actually
arises.

We shall show in this Letter that the missing information is the
chemical and electronic structure of the tip. To this end we
simulate STM measurements from first principles, employing
magnetic and and non-magnetic model tips and the STM code recently
introduced (bSCAN \cite{hofer00a}). In particular we analyze,
whether (i) the corrugation is really substantially enhanced if a
tungsten terminated tip is used \cite{heinze00}; and (ii) whether
contamination of the tip with surface atoms might change the
images. The first point seems important because extensive
simulations of STM scans on an iron surface brought no evidence
for an actual enhancement of the corrugation \cite{hofer00b}. In
addition, it was shown on metal alloy surfaces that a tungsten
terminated tip in fact diminishes rather than enhances the
corrugation \cite{hofer00a}. Furthermore, we could demonstrate
that relaxation effects in the close distance range enhance the
corrugation by about 10 to 15 pm \cite{hofer01a}. Given this
evidence the only route for a fuller understanding are detailed
simulations, which we present in this paper.

The paper is structured as follows. First we present results of
our calculation of the W(110) surface with a single Mn overlayer.
The corrugation of the surface, if evaluated from the local
density of states, will be shown to be very small and in fact
below the limit of STM resolution. Then we demonstrate that due to
the influence of the electronic properties of a tungsten tip the
corrugation is {\it reversed and enhanced} by a factor of about
five. This contrast reversal due to tungsten tips is quite common
on magnetic surfaces \cite{hofer00b}. It is the first time,
however, that we also obtain a substantial contrast enhancement.
Finally, we present the simulated images of the W(110)Mn AFM
surface for an Fe terminated and a Mn terminated tip. Here we show
that the tip most likely used in one of the experiments was
contaminated by Mn atoms.


The electronic structure of the sample and the tip surface were
calculated within density functional theory in a two step
procedure, chosen for efficiency. First, we calculated the
groundstate interlayer distances of the surface and a W(100)X
terminated tip, where X is the single apex atom. For this part of
the calculation we used the Vienna ab initio simulation program
(VASP) \cite{kresse93,kresse96}. This groundstate geometry was
then used in the full-potential augmented plane wave calculation
of electronic surface properties (FLEUR, contrary to the version
described in \cite{kurz00} our code does not include spin-orbit
coupling). This full-potential method is still the most precise
numerical tool for a complex magnetic system. For both steps we
employed a spin-polarized generalized gradient approximation to
exchange and correlation energy according to Perdew et al. (PW91)
\cite{perdew91}.

The STM currents and corrugation were computed with a perturbation
approach based on Bardeen's integral \cite{bardeen61}. The program
computes the integral numerically by summing up all states within
a given energy interval. The interval chosen in our simulations
was 50meV, it is larger than the thermal broadening of states at
the temperature of measurements, which Heinze et al give as 16K
\cite{heinze00}. The reason for this increase is that we sought to
avoid an unrealistically selective evaluation, which might be
distorted by numerical limitations.


We started our simulations by computing the groundstate properties of
the W(110) surface with a FM or AFM Mn adlayer. This adlayer grows
pseudomorphically on the tungsten surface \cite{bode99}. Initially,
we determined the equilibrium position of the ion cores. The
relaxations were performed with VASP. The energy cutoff of the
optimized pseudopotentials \cite{vanderbilt90} for W and Mn was set
to $E_{cut} = 230$eV. We used a Monkhorst-Pack grid
\cite{monkhorst76} of (8 $\times$ 8 $\times$ 1) $k$-points for the
Brillouin-zone integration. The system was simulated by a repeated
slab of seven W layers and one Mn layers on both sides of the
tungsten slab. The vacuum layer separating the repeated slabs was
approximately nine \AA \, thick, this distance is in every case
sufficient to guarantee that adjacent slabs are completely decoupled.
The lattice constant of the tungsten lattice had previously been
computed by minimizing the crystal energy within the PW91 functional
\cite{hofer00a}. There, we obtained a slightly larger constant of
3.19 \AA \, compared to the experimental value of 3.16 \AA. The
interlayer distance between the tungsten surface and the Mn adlayer
is smaller than the ideal bulk interlayer distance, we obtain an
inward relaxation of 5.3 \% (interlayer distance of 2.13 \AA \,
versus 2.25 \AA \, of the ideal lattice). We considered two
configurations: the FM surface and a c(2 $\times$ 2) AFM surface. We
could detect no surface buckling due to the magnetic properties of
the surface atoms. In both cases, the FM as well as the AFM surface,
we obtain the same vertical positions for the two Mn atoms.

The equilibrium positions were the input for the calculation of
electronic groundstate properties using a full potential code
(FLEUR). In this calculation we used the same number of k-points
(16), and the same GGA exchange correlation potential (PW91). Heinze
et al. \cite{heinze00} obtained the AFM surface as the magnetic
groundstate. In our calculation the energy difference between FM and
AFM ordering of -5 meV/Mn is below the level of precision even of
full-potential methods. Therefore we would conclude that both states
are equally probable. However, our calculation did not include
spin-orbit coupling. Spin-orbit coupling on this surface favors AFM
ordering \cite{heinze00} and thus shifts the balance slightly towards
the AFM configuration.

\begin{figure}
\begin{center}
\epsfxsize=1.0\hsize \epsfbox{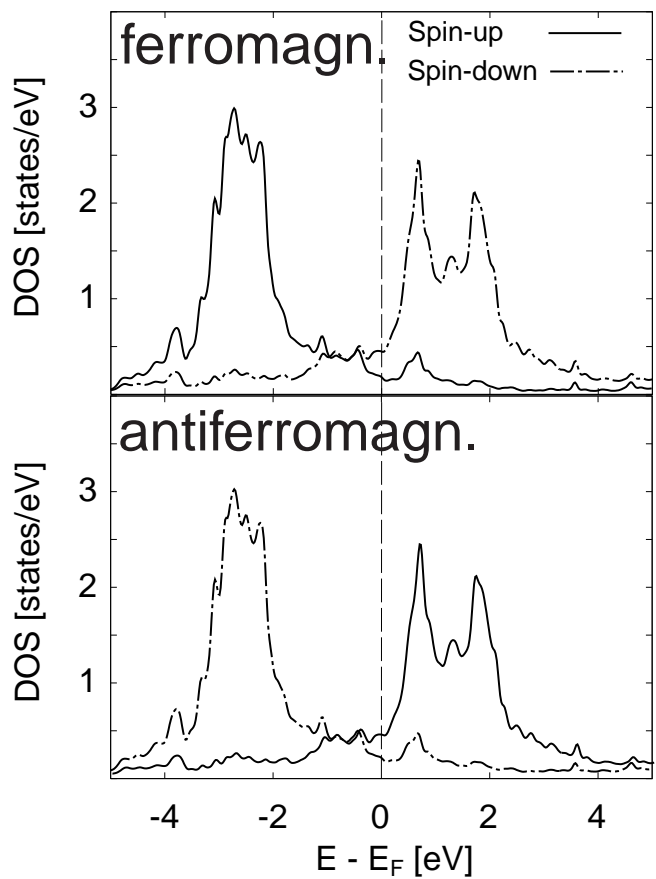}
\end{center}
\caption{
         Density of states (DOS) integrated over the Mn surface
         atoms. The DOS of the ferromagnetic (top) and
         antiferromagnetic (bottom) Mn  atom has the same
         structure, the polarization of spin states is
         reversed.
         }
\label{fig001}
\end{figure}

The induced moment in surface atoms is 0.24 $\mu_{B}$, it decays
into the bulk and is close to zero for the central layer. The
magnetic moment of the Mn atoms is 3.58 $\mu_{B}$, which is a
consequence of the higher interatomic distance in our calculation
due to gradient corrections for the exchange correlation
functional. For the FM surface we obtain a magnetic moment of 3.62
$\mu_{B}$. To evaluate the density of states (DOS) in the Mn layer
we used a triangular mesh of 36 k-points. The DOS of the two atoms
in the AFM layer is shown in Fig. \ref{fig001}. The majority and
minority bands are below and above the Fermi level, respectively,
the spin splitting in our calculation is about 4 eV (middle of
majority to middle of minority band). Even though the exchange
correlation and the lattice constant differ in our calculation,
the DOS is virtually identical to the one obtained by Heinze et
al. \cite{heinze00}. In case of spin-polarized surfaces the
contributions from spin-up and spin-down states are computed
separately. The spin of a tunnelling electron is conserved, which
means that we do not consider the possibility of spin-flips during
tunnelling.

\begin{figure}
\begin{center}
\epsfxsize=1.0\hsize \epsfbox{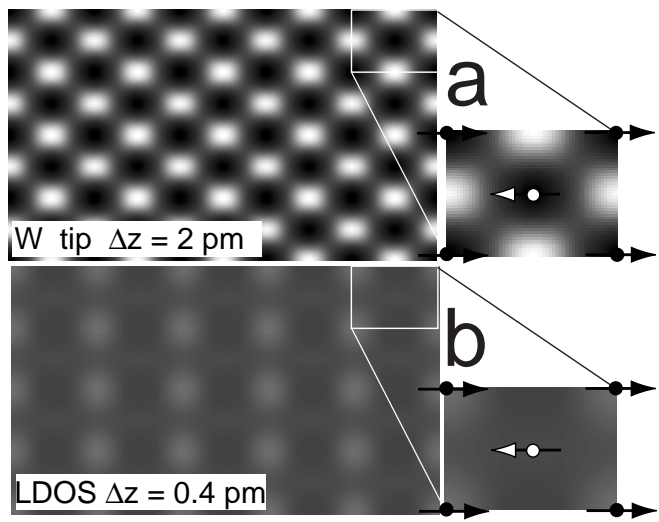}
\end{center}
\caption{
          Corrugation of the surface from a constant current contour
          (a) and from the local density of states (LDOS, (b)). The
          corrugation vanishes for the LDOS. The corrugation with a
          paramagnetic tungsten tip is about 2 pm (tunneling
          conditions: - 3 mV bias and 0.5 nA tunnel current). Note
          that the highest protrusion of the current contour is at
          the hollow site of the W(110)Mn surface. Ferromagnetic
          (full arrow) and antiferromagnetic (empty arrow) surface
          atoms are not discriminated.
         }
\label{fig002}
\end{figure}

Here we present only the results of our simulations for the AFM
surface. Initially, we simulated a scan with a clean tungsten tip.
The electronic structure of this tip is characterized by a high
contribution of $d_{xz}$ and $d_{yz}$ states \cite{hofer00c}.
Furthermore, the DOS in the vacuum has its maximum at + 0.7 eV,
and it is rather low at the Fermi level (0.42 states/eV). For
these reasons a clean tip does not generally produce images of
high contrast. In our simulation this is reflected by the
corrugation height of
2 pm (distance 4.5 \AA, - 3 mV sample bias and 0.5 nA tunnelling
current), shown in Fig. \ref{fig002} (a). The corrugation is lower
than the experimental values by about 13 pm. Considering that the
separation between tip and sample is already at the lower limit of
stability \cite{hofer01a}, we attribute this difference to atomic
relaxations of the system. The paramagnetic tip does not
differentiate between the two Mn atoms; both are equal in the
experiments \cite{heinze00} and the simulations. A paramagnetic
tip thus shows only the chemical structure of the surface. In this
context it seems interesting to differentiate between effects of
the surface electronic structure and effects due to the overlap
between tip and sample wavefunctions. For this reason we have also
plotted the local density of states (LDOS) at the same distance
(see Fig. \ref{fig002} (b)). The corrugation in this simulation is
below the resolution of even a high-precision STM \cite{hofer00b}.
Neither the atomic positions, nor the magnetic structure of the
surface as it appears in the scans can be accounted for by the
electronic properties of the surface alone.

\begin{figure}
\begin{center}
\epsfxsize=1.0\hsize \epsfbox{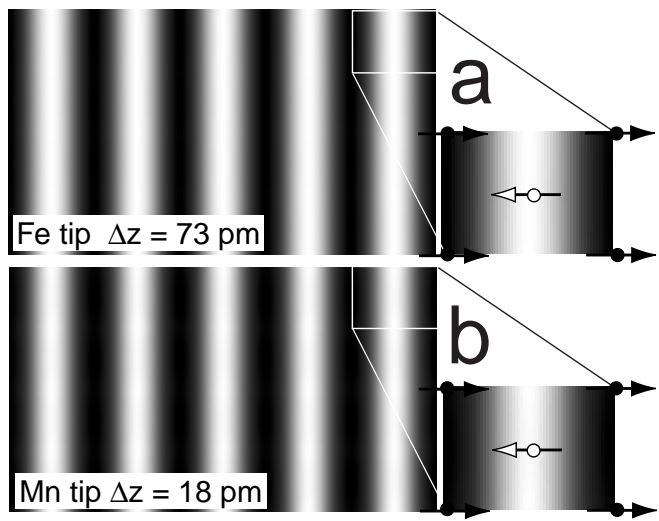}
\end{center}
\caption{
          Corrugation of the surface from constant current contours
          and two different magnetic STM tips. The Fe terminated tip
          resolves the surface magnetic structure with a corrugation
          of 73 pm (a). The Mn terminated tip reveals an identical
          structure but a substantially lower corrugation of 18 pm
          (b). In both cases the tip-sample separation is 4.5 \AA.
          }
\label{fig003}
\end{figure}

The experiments with a ferromagnetic tip used a Fe-coated tungsten
tip. We model the tip by a tungsten film with a Fe apex atom.
Since the chemical nature of the apex atom is commonly the most
important parameter in scans on metals \cite{hofer00a}, a single
Fe atom seems sufficient. It is unclear, however, whether the tip
was contaminated in the experiments by a contact with the Mn
adlayers. Therefore we also account for this possibility by
simulating scans with a Mn terminated tip. The results of our
simulations are presented in Fig. \ref{fig003}. It should be noted
that given plots refer to a median distance of 4.5 \AA. The
current value was chosen accordingly and is smaller than for the
tungsten tip (Fe-tip 0.1 nA, Mn-tip 0.2 nA). The reason for
simulating scans at a specific height rather than a specific
tunnelling current is the absolute value of the current obtained
in the measurements (34 nA). In a simulation this high current
leads to a tip-sample separation well below the level, where the
system is mechanically stable \cite{hofer01a}. For this reason we
estimate the signatures of different tips at a given tip-sample
separation, which we know to be at the lower end of tip stability.

The simulations with magnetic tips in every case reveal the
magnetic structure of the surface. This is due to the changed
coupling of spin-up and spin-down states of sample and tip. A
detailed account of the electronic structure of Fe-terminated and
Mn-terminated tips is given in \cite{hofer01b}. Fig. \ref{fig003}
(a) shows the simulation with a Fe-terminated tip. While the
qualitative features of the simulation are equal to the
experimental results \cite{heinze00}, the corrugation height is
much higher than the experimental value. Judging from results on
other metal surfaces, this feature would be quite unusual. In
fact, a corrugation this much higher than the simulated value has
so far not been observed. For this reason we also simulated scans
with a Mn-terminated tip. This simulation is presented in Fig
\ref{fig003} (b). The qualitative features remain the same, but
the corrugation height (18 pm) is now in accordance with
experimental values, if we assume a partial depolarization of
tunneling electrons due to the orientation of the magnetic axes. A
similar assumption, with a similar result (polarization of tip
states 40\%, corrugation height 7 pm), was made in the previous
paper by Heinze et al. \cite{heinze00}. We therefore conclude that
the actual tip in the measurements was contaminated by Mn atoms of
the surface.

It might seem hard to understand, why the actual current in the
experiments is so much higher than in the simulations (34 nA
versus 0.1 to 0.5 nA). In the simulations we obtained the current
values by assuming that the distance between tip and sample is at
the lowest limit of mechanical stability, or about 4.5 \AA
(core-core distance) \cite{hofer01a}. Generally, the tunnel
current on a metal surface under these conditions amounts to 1-10
nA in simulations \cite{hofer01a}. That it is only one tenth of
this value in the present situation is due to the low density of
states at the Fermi level (see Fig. \ref{fig001}). This low
density of states yields a number density at the vacuum boundary
of $n({\bf r}) \approx 2 \times 10^{-3}$\AA$^{-3}$, which is about
10\% of the number density encountered at the vacuum boundary of
transition metals \cite{pettifor95}. Note that the same reduction
due to the low density is encountered by every method of STM
simulation, whether it is based on scattering theory or
perturbation theory, as long as it relies on density functional
theory.

Within groundstate density functional theory only two effects
could lead to substantially changed current values: (i) The change
of the magnetic splitting due to spin-orbit coupling. (ii) The
change of the exponential decay due to image potentials. Both
effects seem negligible in the present situation. The first,
because the density of states is the same, whether spin orbit
coupling is neglected, as in our calculation, or whether it is
included, as in the previous calculation \cite{heinze00}. The
second does not change the number density at the vacuum boundary,
which is already too low for any agreement with experimentally
measured currents. So that we arrive at the conclusion that no
current treatment of STM, based on first principles density
functional theory, is suitable to reproduce the measured current
values.

It is conceivable that the difference is due to excitations or
dynamic processes during tunneling. The energy range, for example,
of tunneling electrons could not be limited to the Fermi level, if
the system is excited. This question has been analyzed
theoretically by Todorov, and it was found that the temperature
gain due to a bias voltage in the range of a few mV is about 100K
\cite{todorov98}. For a substantial change the energy of the
tunneling electrons would have to be about 2eV below the Fermi
level. Since this energy shift corresponds to a temperature change
of more than 1000K, we can exclude this possibility. Finally, it
seems possible that phonon excitations due to interactions between
tip and sample have an effect on the tunneling current. Here, we
find that the change of the tunnel current due to phonon
excitations is less than 10\% even for highly polar molecular
bonds \cite{persson00}. In summary, we can exclude that any of
these effects may alter the tunnel current by two orders of
magnitude. Therefore the disagreement between experiments and
simulations concerning the measured current values cannot be
removed in any treatment relying on groundstate density functional
theory. Since this theory has been so effective in the qualitative
and quantitative modelling of experiments in condensed matter
physics, we think that the disagreement provides a good reason for
a careful revision of the experiments. Considering the acceptable
agreement between measured and calculated corrugations and pending
further experimental results we think that the main contributions
to the measured current should be independent of the lateral
position of the STM tip, i.e. due to a current background not
covered in our simulation.

The work was supported by the British Council and the National
Research Council. Computing facilities at the UCL HiPerSPACE
center were funded by the Higher Education Funding Council for
England. AJF was also supported by an Advanced Fellowship from the
Engineering and Physical Sciences Research Council. Helpful
discussions with Matthias Bode and Stefan Bl\"ugel are gratefully
acknowledged.

%

%

\end{document}